\documentclass[preprint]{emulateapj}

\usepackage{aas_macros}

\usepackage[dvips]{color}

\slugcomment{To Appear in ApJ}
\shortauthors{Chiaki, Yoshida, \& Kitayama}
\shorttitle{Low-metallicity star formation}

\begin{document}

\title{Low-mass star formation triggered by early supernova explosions}

\author{
Gen Chiaki$^{\dag}$\altaffilmark{1},
Naoki Yoshida$^{\dag}$
and
Tetsu Kitayama$^{\ddag}$
} 
\affil{
$^{\dag}$Department of Physics, Graduate School of Science, University of Tokyo, 
7-3-1 Hongo, Bunkyo, Tokyo 113-0033, Japan\\
$^{\ddag}$Department of Physics, Toho University, Funabashi, Chiba 274-8510, Japan
}
\altaffiltext{1}{E-mail: gen.chiaki@utap.phys.s.u-tokyo.ac.jp}

\begin{abstract}
We study the formation of low-mass and extremely metal-poor stars
in the early universe. Our study is motivated by the recent 
discovery of a low-mass ($M_*\leq 0.8 \ M _{\odot}$) and extremely 
metal-poor ($Z \leq 4.5 \times 10^{-5} \ Z_{\odot}$) star 
in the Galactic halo by \citet{Caffau11}.
We propose a model that early supernova (SN) explosions 
trigger the formation of low-mass stars via shell fragmentation.
We first perform one-dimensional hydrodynamic simulations of the evolution of 
an early SN remnant. We show that the shocked shell undergoes 
efficient radiative cooling and then becomes gravitationally 
unstable to fragment and collapse in about a million years.
We then follow the thermal evolution of the collapsing fragments
using a one-zone code. Our one-zone calculation treats chemistry 
and radiative cooling self-consistently in low-metallicity gas.
The collapsing gas cloud evolves roughly isothermally, 
until it cools rapidly by dust continuum emission
at the density
$10^{13}$--$10^{14} \ {\rm cm}^{-3}$. 
The cloud core then becomes unstable and fragments again.
We argue that early SNe can trigger the formation of
low-mass stars in the extremely metal-poor environment
as \citet{Caffau11} discovered recently.
\end{abstract}



\begin{figure*}[t]
\epsscale{0.80}
\plottwo{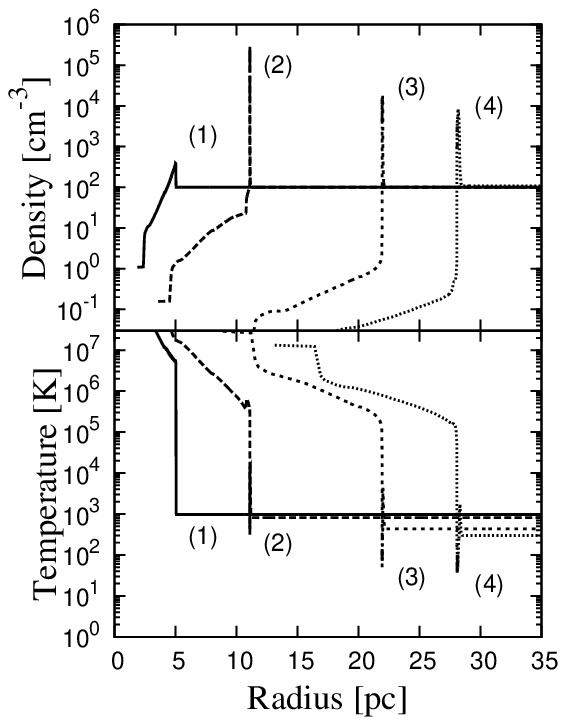}{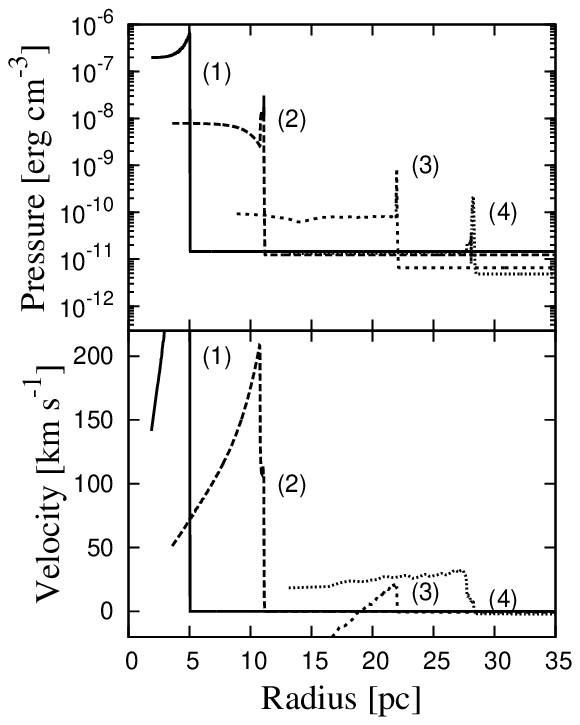}
\caption{
We plot the evolution of 
the density, temperature, pressure, and velocity 
profiles of the supernova remnant 
for $(E_{{\rm SN}},n_0,Z)
=(1\times 10^{52} \ {\rm erg} ,100 \ {\rm cm^{-3}}, 1.0\times 10^{-5} \ Z_{\odot })$ 
at the time
(1) $t=3.2\times 10^{3}$, 
(2) $3.2\times 10^{4}$, 
(3) $3.2\times 10^{5}$, and
(4) $8.6\times 10^{5}$ yr.
The blast wave driven by SN explosion
sweeps up the ambient medium and
the SN-shocked shell is formed.
Epoch (1) corresponds to the Sedov-Taylor phase,
in which the shell expands adiabatically.
Then, the cold dense shell forms by radiative cooling
at epoch (2), and eventually
the shell is decelerated sufficiently to
be gravitationally unstable at epoch (4).
}
\label{fig:prof_L}
\end{figure*}

\section{INTRODUCTION}

Stars with masses $M_* \lesssim 0.8 M_{\odot }$ 
formed in the early universe may survive to the present day.
Such old low-mass stars should typically have very low metallicites
because they were formed from a gas that had not been
significantly enriched with heavy elements. 
They carry invaluable information 
on the early chemical evolution of the Galaxy \citep{Beers05}.
Recently, \citet{Caffau11} discovered 
a low-mass and extremely metal-deficient star
in the halo of the Galaxy.
The star has extremely small abundances of carbon and nitrogen.
The upper limit of corresponding metallicity of the star is
estimated to be $4.5 \times 10^{-5} \ Z_{\odot}$ \citep{Caffau11}.
Furthermore, the ratio of the abundances of carbon and 
oxygen derived from the spectra of the star
suggests that the parent cloud from which the star
was born had been chemically enriched by 
core-collapse supernovae
\citep{Caffau11}.

It is generally thought that metal-free primordial stars
are predominantly massive \citep[{\it e.g.},][]{Bromm09}
although not extremely massive \citep{Hosokawa11}.
This is largely because metal-free gas lacks
efficient coolants other than hydrogen molecules; 
a primordial gas cloud is thermally stable throughout
its pre-stellar collapse 
\citep{OmukaiYoshii03, Yoshida08}.
Theoretical studies suggest that
a trace amount of heavy elements 
provide efficient cooling in low-metallicity
gas, which may then enable the formation of
low-mass stars \citep[see][for a review]{BrommYoshida11}.
There is another argument that the existence of
dust grains in low-metallicity gas is crucial
for the low-mass star formation
\citep{Schneider03,Omukai05,OmukaiHosokawaYoshida10}.
Although it is still unclear 
whether or not gas 
metallicity is the key quantity to determine
the characteristic stellar mass, 
and whether or not there is ``critical metallicity'' 
for low-mass star formation, 
there must be a mechanism for the formation of low-mass
stars in extremely low-metallicity gas\footnote{
In this paper, we use the term ``low-mass'' for a mass 
less than $1 \ M_{\odot }$}.

SN explosions of the first generation stars are thought to have enriched 
the early universe with the first heavy elements.
Three-dimensional simulations of early SN explosions are
performed by, {\it e.g.}, \citet{Bromm03} and \citet{Greif10}.
These studies show heavy elements are effectively dispersed
into the inter-galactic medium.
\citet{Whalen10} argue that ultraviolet radiation from a massive progenitor 
star modifies the density structure of the ambient gas before SN explosion.
Overall, an early SN explosion can trigger the formation of low-mass and low-metallicity stars.
\citet{Nagakura09} study the evolution of SN remnants with metallicities 
of $Z=10^{-4}$ and $10^{-2}\ Z_{\odot }$.
The SN shell becomes gravitationally unstable for a certain range of parameters 
such as explosion energy, ambient gas density, and metallicity.
They show that the typical mass scale of the fragments is $\sim 10^3$--$10^4 \ M_{\odot}$
and argue that a low-mass star forms after the further evolution of the fragment.
\citet{Machida05} semi-analytically calculate the evolution of SN shells 
before and after the shell fragmentation in a primordial gas. They show 
that the Jeans mass of the fragment is below the solar mass when
the fragment becomes optically thick. However, it remains uncertain whether 
the gas cloud further fragment or not.

In this paper, we study SN-triggered star formation,
using hydrodynamic simulations.
We first run one-dimensional
hydrodynamic simulations to follow the thermal evolution
of an early SN remnant.
We here assume the ambient gas around the SN
has nearly the primordial composition with a trace amount of heavy
elements, $Z = 1.0\times 10^{-5} Z_{\odot}$ and 
$4.5\times 10^{-5} Z_{\odot}$, presumably ejected by earlier
generation stars. 
We determine whether or not the shocked SN shell is
gravitationally unstable to fragment 
by using linear analysis of density perturbations.  
We show that gravitational
instability is triggered in the shell 
for a wide range of the supernova
explosion energies, $E_{{\rm SN}} = 1\times 10^{51}$--$3\times 10^{52}$ erg,
and for large ambient gas density, $n_0\geq 10 \ {\rm cm^{-3}}$.
Finally, we study the subsequent evolution of the collapsing fragments,
using a one-zone code.  
While the gas cloud continues collapsing gravitationally, it cools rapidly 
by dust thermal emission and eventually fragments to sub-solar mass clumps.


\section{NUMERICAL METHOD}

\subsection{Code}

\subsubsection{Ante-fragmentation phase}

We perform one-dimensional hydrodynamic calculations
to follow the dynamics of an SN remnant until the shocked gas shell 
undergoes gravitational instability 
(hereafter, we call this phase 
{\it ante-fragmentation phase}).
We use the code of \citet{Kitayama04} 
and \citet{KitayamaYoshida05}
which employs the second-order 
Lagrangian finite-difference scheme
in spherically symmetric geometries.
The basic equations are presented in 
$\S$2 of \citet{Kitayama04}.

We implement the relevant radiative cooling processes 
to the hydrodynamical code as follows.
In the high temperature regime ($T>10^4 \ {\rm K}$), assuming
that the gas is in collisional ionization equilibrium, we adopt 
the cooling function of \citet{SutherlandDopita93}.  
The cooling functions for $Z<10^{-3} \ Z_{\odot }$
are derived by interpolating the cooling functions 
for $Z=10^{-3} \ Z_{\odot }$ and $Z=0$. 
The cooling rate due to inverse Compton scattering is also included.

In the lower temperature regime ($T<10^4 \ {\rm K}$), 
we solve the non-equilibrium chemical reactions 
and radiative processes of
the following fifteen species: $e^-$, H, H$^+$, H$^-$, D, D$^+$, D$^-$,
He, He$^+$, He$^-$, H$_2$, H$_2^+$, HD, HD$^+$, and HeH$^+$.
The molecular cooling rates are taken
from \citet{GalliPalla98} for H$_2$, \citet{Lipovka05} 
and from \citet{Flower00} for HD.  
We also calculate the fine-structure transition line cooling
of heavy elements including C, O, Si, and Fe.
These species are assumed to be in the form of \ion{C}{2}, \ion{O}{1},
\ion{Si}{2}, and \ion{Fe}{2}, respectively.
The fraction of each species in the gas is derived by multiplying
the solar abundance by metallicity ($Z/Z_{\odot}$).
The solar abundances of C, O, Si, and Fe
relative to hydrogen are $2.45\times 10^{-4}$, 
$4.90\times 10^{-4}$, $3.63\times 10^{-5}$, and $3.16\times 10^{-5}$,
respectively \citep{SantoroShull06} throughout the one-dimensional calculation.
Also, the atomic level transition rates are taken 
from \citet{SantoroShull06}.

\subsubsection{Postfragmentation phase}

After the supernova shell becomes gravitationally unstable
(see \S \ref{subsec:criterion} below),
fragments in a shell are expected to 
collapse further to trigger star formation.
We perform one-zone calculations
to follow the thermal and chemical evolution of one of 
such collapsing gas clouds.
We assume that the collapse occurs approximately in a 
spherically symmetric manner.
Then, the time evolution of the gas density, $\rho$, 
is described as
\begin{equation}
\frac{d\rho}{dt} = \frac{\rho}{t_{{\rm ff}}}.
\label{one-zone:drhodt}
\end{equation}
In this equation, $t_{{\rm ff}}$ denotes the free-fall time as
\begin{equation}
t_{{\rm ff}}=\left( \frac{3\pi}{32G\rho} \right)^{1/2},
\label{eq:tff}
\end{equation}
where $G$ is the Newtonian gravitational constant.
The evolution of specific internal energy, $e$, is
\begin{equation}
\frac{de}{dt}= -p \frac{d}{dt} 
\left(
  \frac{1}{\rho}
\right)
 + 
\frac{\Gamma - \Lambda}{\rho },
\label{eq:dudt}
\end{equation}
where 
$p$ denotes the pressure and
$\Gamma$ and $\Lambda$ denote heating and cooling rate
per unit volume, respectively.
The temperature of the cloud, $T$, is related to the internal energy
as
\begin{equation}
e=\frac{1}{\gamma -1}
\frac{kT}{\mu m_{{\rm H}}},
\label{eq:u2T}
\end{equation}
where $\mu$ is the mean molecular weight,
$m_{{\rm H}}$ is the mass of hydrogen nuclei.
In equation (\ref{eq:u2T}), $\gamma $ is the adiabatic index 
which is calculated by taking
average over all species as {\it e.g.}, \citet{OmukaiNishi98}.

In our one-zone model, the low-metallicity gas is 
assumed to consist of the following chemical species:
$e^-$,
H, H$^+$, H$^-$, D, D$^+$, D$^-$,
He, He$^+$, He$^-$,
C, C$^+$, O, O$^+$,
H$_2$, H$^-_2$, H$^+_2$,
HD, HD$^+$, HD$^-$,
OH, 
H$_2$O,H$_2$O$^+$,
O$_2$, O$_2^+$,
CO, CO$^+$, CO$_2$, and CO$_2^+$.
We assume that 72 \% of carbon and 46 \% of oxygen 
are condensed to dust grains
as in the local interstellar matter \citep{Pollack94}.
We include atomic cooling by 
\ion{C}{2} and \ion{O}{1}
but neglect the contributions from
less abundant \ion{Si}{2} and \ion{Fe}{2}
because these species do not have effect on
the thermal evolution of the gas cloud
\citep{Schneider12}.
The adiabatic compressional heating and
the chemical heating by H$_2$ formation
are the dominant heating processes.
We also consider the dust cooling in the calculations.
The reaction rates and corresponding cooling rates are
taken from \citet{Omukai05,OmukaiHosokawaYoshida10}.
We use the Sobolev length approximation for optically thick cooling
as in \citet{Omukai00} and \citet{Yoshida06}.


\begin{figure}[t]
\epsscale{1.0}
\plotone{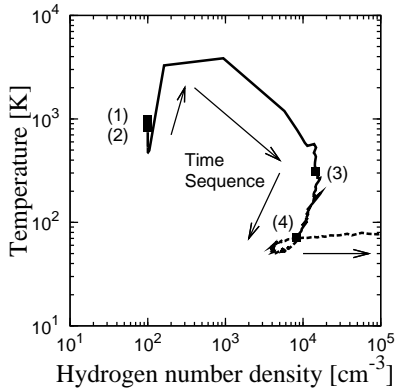}
\caption{
We plot
the evolution of the density and temperature 
of a fluid element, which is initially located
at 20.8 pc from the center.
Here, the result for the SN model of
$(E_{{\rm SN}},n_0,Z)
=(1\times 10^{52} \ {\rm erg} ,100 \ {\rm cm^{-3}}, 1.0\times 10^{-5} \ Z_{\odot })$ 
is shown.
Solid and dashed curves denote the evolution before and 
after fragmentation, respectively.
Both results are obtained by our one-dimensional calculations.
Points (1) to (4) indicate the epochs
in Figure \ref{fig:prof_L}.
After epoch (4), the density in the region 
turns to increase by self-gravity.
}
\label{fig:nT1D_L}
\end{figure}

\subsection{Initial Conditions}

We set metallicities of the circumstellar medium
$Z=1.0 \times 10^{-5} \ Z_{\odot}$ conservatively
and $4.5 \times 10^{-5} \ Z_{\odot}$, the upper bound
of the metallicity of the star \citet{Caffau11} observed.
We assume that the progenitor star is hosted by an early
small dark matter halo with mass $M_{{\rm h}}=10^6 \ M_{\odot}$
at a redshift of $z=20$ \citep{Yoshida03}.
The gravitational potential of the dark matter is 
fixed as in \citet{KitayamaYoshida05}.
We work with the $\Lambda$CDM model with
the matter density $\Omega _{{\rm m}}=0.27$, 
the baryon density $\Omega _{{\rm b}}=0.045$,
the cosmological constant $\Omega _{\Lambda}=0.73$, 
the Hubble constant $h=0.71$,
in units of $100 \ {\rm km \ s^{-1} \ Mpc^{-1}}$,
and the absolute density fluctuation $\sigma _8 = 0.80$
consistent with WMAP 7 data \citep{Larson11}.

The density of the circumstellar medium at the time when SN explosions
occur depends on a number of factors such as the mass of the hosting halo 
\citep{Whalen04, Whalen08, Kitayama04} and the ultraviolet luminosity of the 
progenitor star \citep{Shu02,Alvarez06}. Thus, for simplicity,
we assume that the surrounding gas density is initially uniform for our 
one-dimensional hydrodynamic simulations and the initial density is 
set $n_0 =$1, 10, 100, and 1000 ${\rm cm^{-3}}$.

We assume that the surrounding gas with primordial composition is 
pre-enriched with a trace amount of heavy elements.
Thus, through this paper, the hydrogen mass fraction is $X=0.74$, 
the helium mass fraction is $Y=0.26$
\citep{Aver10},
and the deuterium number fraction is 
${\rm D}/{\rm H}=2.9\times 10^{-5}$
\citep{Iocco09}.
We set the initial gas temperature $T=10^3 \ {\rm K}$, and
initial ionization fraction of hydrogen is uniformly
$10^{-4}$. The fraction of hydrogen molecule is 
$10^{-3}$ at the beginning\footnote{
The structure and evolution of SN shells are indeed insensitive to the 
circumstellar temperature and chemical compositions because the upstream 
pressure is negligible against the downstream pressure at the beginning 
of the shell evolution.
We also confirm that at the end of simulations the upstream temperature 
becomes below $10^3$ K by radiative cooling even if we start the simulation 
with the temperature of $10^4$ K.}.
The radial velocity is set $v=0 \ {\rm km \ s^{-1}}$.
We test three explosion energies, $E_{{\rm SN}} = $ 
$1 \times 10^{51}$,
$1 \times 10^{52}$, and 
$3 \times 10^{52}$ erg.
We assume a point-source explosion and 
at the time $t=0$, the corresponding thermal energy is deposited
in the innermost cell.


\begin{figure*}
\epsscale{0.80}
\plottwo{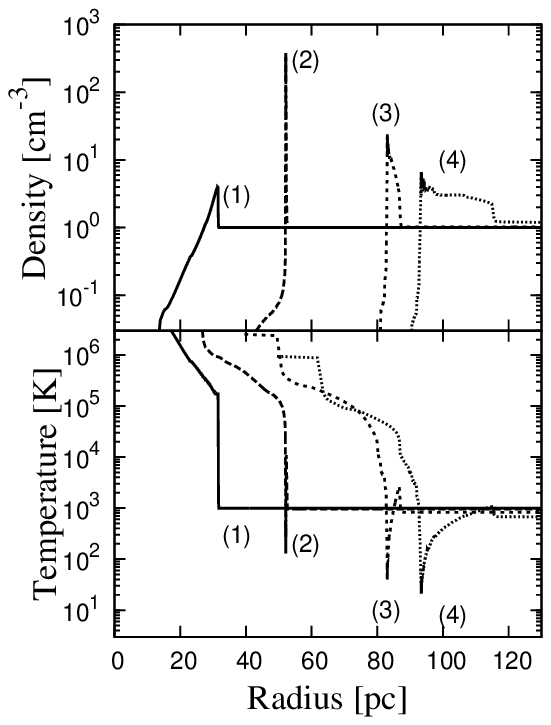}{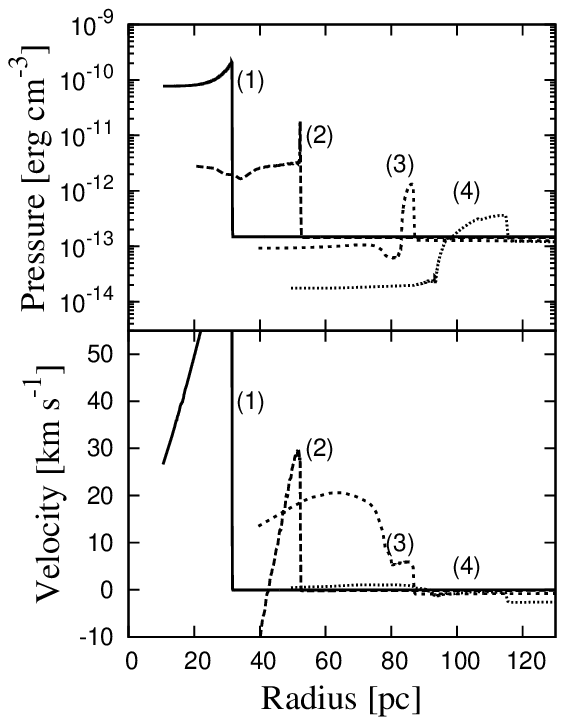}
\caption{
Same as Figure 1, but for
$(E_{{\rm SN}},n_0,Z)
=(1\times 10^{51} \ {\rm erg} ,1 \ {\rm cm^{-3}}, 1.0\times 10^{-5} \ Z_{\odot })$ 
at the time
(1) $t=1.0\times 10^{5}$, 
(2) $5.6\times 10^{5}$, 
(3) $3.2\times 10^{6}$, and
(4) $1.0\times 10^{7}$ yr.
}
\label{fig:prof_S}
\end{figure*}

\section{RESULTS}

\subsection{Shell Evolution}
\label{subsec:ShellEvo}

\subsubsection{Larger ambient gas density}

Figure \ref{fig:prof_L} shows the structures of the shell
for $(E_{{\rm SN}},n_0,Z)
=(1\times 10^{52} \ {\rm erg} ,100 \ {\rm cm^{-3}}, 1.0\times 10^{-5} \ Z_{\odot })$ 
at the time
(1) $t=3.2\times 10^{3}$, 
(2) $3.2\times 10^{4}$, 
(3) $3.2\times 10^{5}$, and
(4) $8.6\times 10^{5}$ yr.
Figure \ref{fig:nT1D_L} illustrates the evolution 
of the gas density and temperature for a
fluid element that is densest when the shell becomes gravitationally unstable 
and that is initially located at 20.8 pc from the center.
The evolution of the SN shell structure is described as follows.
In the initial Sedov-Taylor phase, 
the SN remnant expands adiabatically as long
as the radiative cooling time, $t_{{\rm cool}}=\rho e/\Lambda $,
is longer than 
the shell expansion time scale, $t_{{\rm exp}} = R/V$
\citep{Sedov58}.
Then, radiative cooling becomes effective
and the gas just behind the shock starts to
lose its thermal energy.
At $t\simeq 2\times 10^4$ yr after
the SN explosion,
the cold shell is pushed by the large 
pressure of the inner hot gas,
forming a very dense shell
(the pressure-driven snowplough phase).
During the pressure-driven phase,
the fluid element plotted in Figure \ref{fig:nT1D_L}
enters the shock front, increasing its density and 
temperature rapidly.
The inner hot region still impacts the cold shell, 
generating multiple weak shocks behind the shell. 
We see this as oscillatory features from epoch (3) to epoch (4)
in Figure \ref{fig:nT1D_L}.
Finally, at epoch (4), the shell is decelerated
sufficiently to be gravitationally unstable
as we shall discuss in \S \ref{subsec:criterion}.

\subsubsection{Smaller ambient gas density}
\label{subsubsec:S}

\begin{figure}
\epsscale{1.0}
\plotone{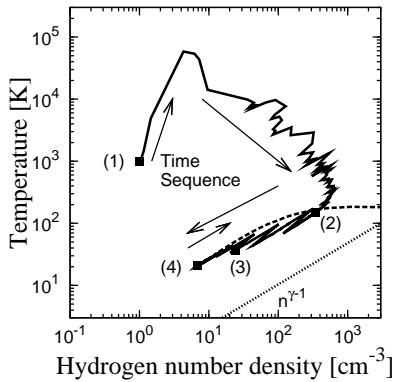}
\caption{
Same as Figure 2, but for
$(E_{{\rm SN}},n_0,Z)
=(1\times 10^{51} \ {\rm erg} ,1 \ {\rm cm^{-3}}, 1.0\times 10^{-5} \ Z_{\odot })$ .
Points (1) to (4) indicate the epochs
in Figure \ref{fig:prof_S}.
Dotted line shows an adiabatic trajectory,
$T \propto n^{\gamma -1}$.
The trajectory is along an adiabatic line between epochs (2) and (4)
because the shell expands radially for small external pressure, while in larger
ambient density case, the density declines slowly because
the shell expansion is suppressed by the large external pressure and the temperature
decreases by radiative cooling (Figure \ref{fig:nT1D_L}).
}
\label{fig:nT1D_S}
\end{figure}

Figure \ref{fig:prof_S} shows the structures of the shell
for $(E_{{\rm SN}},n_0,Z)
=(1\times 10^{51} \ {\rm erg} ,1 \ {\rm cm^{-3}}, 1.0\times 10^{-5} \ Z_{\odot })$ 
at the time
(1) $t=1.0\times 10^{5}$, 
(2) $5.6\times 10^{5}$, 
(3) $3.2\times 10^{6}$, and
(4) $1.0\times 10^{7}$ yr.
Figure \ref{fig:nT1D_S} illustrates the evolution 
of the gas density and temperature for a
fluid element initially located at 34.3 pc from the center.  
In this model, the shell is in the Sedov-Taylor phase around epoch (1)
and then enters the pressure-driven snowplough phase at $t \simeq 2\times 10^5$ yr.
In this case with low ambient gas density, the transition from 
the Sedov-Taylor phase to the pressure-driven phase occurs later than the model with 
high ambient gas density.
When the shell sweeps up denser ambient medium,
radiative cooling of the shell,
which is generally proportional to the square of gas density, 
becomes larger, and the enhanced radiative cooling makes
the shell evolve faster.
The shell continues to expand even after 
the pressure inside the shell declines due to radiative
cooling, conserving the momentum
(the momentum-conserving snowplough phase).
In our calculation, the shell enters the momentum-conserving phase
by epoch (3).
The shell thickness increases
because of the pressure in the shell. 
The internal energy of the shell remains higher by
less effective radiative cooling.
Around epoch (4), the ellapsed time reaches
the dynamical timescale for the dark matter halo
hosting the progenitor star ($t_{{\rm dyn}} \sim 10^7$ yr for 
$10^6 \ M_{\odot }$ halo at $z\sim 20$).

\subsection{Shell Fragmentation}
\label{subsec:criterion}

\subsubsection{Criteria for fragmentation}

There are several criteria proposed for the instability 
of an expanding and decelerating gas shell.  
\citet{Elmegreen94} derive approximate conditions for 
the collapse of an expanding and decelerating shell 
by linear analysis of density perturbations
under the thin-shell approximation.
Then, \citet{IwasakiTsuribe08} and \citet{IIT11i,IIT11ii}
include the effect of the internal structure of the shell
to the result by \citet{Elmegreen94}.
For a shell with average column density over the 
all solid angle, $\Sigma _0$, shell radius, $R$, and velocity $V$, 
the angular wavenumber of the perturbations that grow most rapidly is 
\begin{equation}
\eta _{{\rm peak}} = \frac{\pi G \Sigma _0}{c_{{\rm eff}} ^2}R,
\label{eq:eta_peak}
\end{equation}
and the growth rate of the perturbations with wavenumber $\eta _{{\rm peak}}$ is 
\begin{equation}
\omega _{{\rm peak}} 
= -\frac{3V}{R}+\left[ \left( \frac{V}{R} \right) ^2
+ \left( \frac{\pi G \Sigma _0}{c_{{\rm eff}}} \right) ^2 \right] ^{1/2}.
\label{eq:o_peak}
\end{equation}
In equations (\ref{eq:eta_peak}) and (\ref{eq:o_peak}),
$c_{{\rm eff}}$ is the effective sound speed as
\begin{equation}
c_{{\rm eff}}=\left[ A\frac{2\pi G \Sigma _0^2}{\rho_{{\rm max}}}
+ \left(\frac{c_T}{2} \right) ^2 \right] ^{1/2},
\label{eq:c_eff}
\end{equation}
where $\rho _{{\rm max}}$ is the peak density in the shell and
$c_T=(kT/\mu m_{{\rm H}})^{1/2}$ is the sound speed for an 
isothermal shell with temperature $T$.
The numerical factor $A$ is determined as 0.39  
by the detailed linear analysis 
for a shell with its outer boundary defined by the shock front 
and with its inner boundary defined
by the stellar wind from a central star
\citep{IIT11i,IIT11ii}.
Note that replacing $c_{{\rm eff}}$ with $c_T$
and $\Sigma _0$ with $\rho_0 R/3$, where $\rho_0$ is the average
pre-shell mass density, in equations (\ref{eq:eta_peak}) and (\ref{eq:o_peak})
recovers the result of \citet{Elmegreen94}.

In equation (\ref{eq:o_peak}),
the first and second terms in the right hand side
represent the effect of the shell expansion and the last term represents the
effect of the shell internal pressure and self-gravity.
If the peak growth rate, $\omega _{{\rm peak}}$, is positive,
perturbations with the corresponding wavenumber have a growing solution,
which means the SN-driven shell becomes gravitationally unstable
and fragments with scales corresponding to $\eta _{{\rm peak}}$ start to grow in the shell.
It is obvious that fast expansion (large $V$) tends to smooth out perturbations.
For the gas to become gravitationally unstable,
its expansion velocity must be sufficiently small such that 
\begin{equation}
\frac{V}{R} < \frac{\pi G \Sigma _0}{\sqrt{8}c_{{\rm eff}}}.
\label{eq:cri}
\end{equation}

\subsubsection{Application of the criterion to our results}

In this section, we discuss whether 
the gravitational instability can be triggered or not.
We apply the criterion by \citet{IIT11i,IIT11ii}
to our calculations
because the internal structure of a shell 
is taken into account in their analysis\footnote{ 
\citet{IIT11i,IIT11ii}
presented the linear analysis for a shell defined by a shock front
and an inner boundary formed by the pressure of the
inner region. 
It is thus more appropriate to our results of the 
SN-driven shell defined by the shock front and the inner 
boundary although the shell inner boundary in the study of 
\citet{IIT11i,IIT11ii} is driven by the pressure
of the stellar wind from the central star.
We have also checked that the result
is qualitatively unchanged if we apply the result 
by \citet{Elmegreen94} which ignores the effect of
the shell structure.}.
In order to investigate the stability of the 
shell, we calculate the time evolution of the 
peak growth rate, $\omega _{{\rm peak}}$ [equation (\ref{eq:o_peak})]
by using the radius of the shock front, the shock velocity, 
volume-weighted mean density, mass-weighted 
mean temperature, and mean molecular weight in the shell.
In the snowplough phase, the shell is defined by the 
outer boundary (shock front) and the inner boundary.
The outer boundary is determined as position where
the density is 1.1 $\rho _0$.
The displacement of the inner boundary
is defined as the point where density recovers 
the pre-shell density, $\rho _0$, inside the 
shock radius.

After the shell becomes gravitationally unstable,
the shell is assumed to be separated into fragments.
If the time for fragmentation is comparable to or larger
than the dynamical timescale of the hosting halo, $t_{{\rm dyn}}$,
we do not count such cases as successful fragmentation of the
SN shell in the cosmological context.
After the dynamical timescale, it is likely that the halo 
itself collapses or is incorporated into a large halo,
thereby the structure of the entire system would be much different.

Applying the fragmentation condition, we find that
for the SN model of $(E_{{\rm SN}},n_0,Z)
=(1\times 10^{52} \ {\rm erg} ,100 \ {\rm cm^{-3}}, 1.0 \times 10^{-5} \ Z_{\odot })$,
the compressed and decelerated shell becomes 
gravitationally unstable 
at $t_{{\rm inst}}=8.3\times 10^5$ yr.
By contrast, for $(E_{{\rm SN}},n_0,Z)
=(1\times 10^{51} \ {\rm erg} ,1 \ {\rm cm^{-3}}, 1.0 \times 10^{-5} \ Z_{\odot })$,
the peak growth rate $\omega _{{\rm peak}}$ becomes positive
at $t_{{\rm inst }}=1.0\times 10^7$ yr
when the shell is about to fall back ($V\simeq 0$) by the
gravitational potential of the host halo.
In this model, since the time $t_{{\rm inst }}$ is 
comparable to the dynamical timescale,
it is uncertain whether the shell fragments 
or collapses into the center of the halo.

Table \ref{tab:frag} shows the
physical properties of the shell
at the time $t_{{\rm inst}}$
for SN models we study.
The results are shown only for $Z=1.0\times 10^{-5} \ Z_{\odot }$
because they are almost exactly the same
as the results of the $Z=4.5\times 10^{-5} \ Z_{\odot }$ models.
It is because the fine-structure cooling of metals
such as C and O has little effect on the evolution of the SN remnant
in such an extremely metal-poor regime
until the shell becomes gravitationally unstable.
Also, the time $t_{{\rm inst}}$
is insensitive to the SN explosion energy.
For a radiative remnant, the shock radius
increases only as $R\propto E_{{\rm SN}}^{1/7}$
in the pressure-driven phase \citep{OstrikerMcKee88}.
The time $t_{{\rm inst}}$
depends on the ambient gas density.
As the ambient gas density is larger,
denser and colder shell is formed as
we discuss in \S \ref{subsubsec:S}
and the growth rate of the perturbations
on the shell, $\omega_{{\rm peak}} $ becomes positive earlier.
The Jeans mass (typical mass) of the fragment is shown in the last column of Table \ref{tab:frag}.
The mass decreases as ambient gas density increases because the density of the shell swept up by
the blast wave becomes larger.

\begin{figure}[t]
\epsscale{0.80}
\plotone{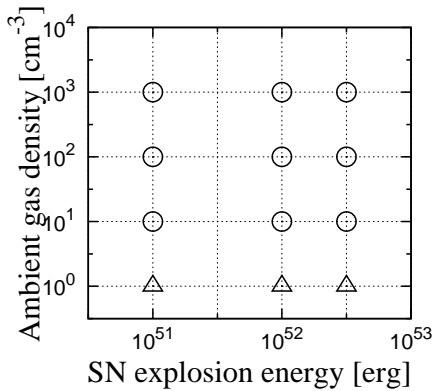}
\caption{
Fragmentation conditions for each our SN model.
The result is common between the model with 
$Z=1.0\times 10^{-5} \ Z_{\odot}$ and
$4.5\times 10^{-5} \ Z_{\odot}$.
The circles indicate models where fragmentation condition
is met. The triangles indicate 
models for which it is uncertain by our calculations whether 
the shell fragments or not (see text).
}
\label{fig:ox}
\end{figure}

Figure \ref{fig:ox} shows the results of all
the models that we investigated in a ($E_{{\rm SN}},
n_0$) plane. 
These results are essentially the same for the respective 
model with $4.5\times 10^{-5} \ Z_{\odot }$.
The circles indicate models where both our criteria are met.
For models with ambient density $n_0 = 1 \ {\rm cm^{-3}}$,
the time for the gravitational instability, $t_{{\rm inst}}$,
is comparable to $t_{{\rm dyn}}$.
These models are marked by triangles.
We cannot determine from our 1D calculations
whether the shell fragments before the host halo evolves significantly.
It would be needed to run a cosmological simulation that incorporate
the hierarchical growth of dark halos, in order to follow the
evolution of such systems properly.

\begin{deluxetable*}{rrrrrrrrrr}
\tablecolumns{10}
\tablecaption{Properties of Fragments
($Z=1.0\times 10^{-5} \ Z_{\odot}$)
\label{tab:frag}}
\tablewidth{0pt}
\tablehead{
\colhead{$n_0$}  &                 
\colhead{$E_{{\rm SN}}$}  &        
\colhead{$t_{{\rm inst}}$} &       
\colhead{$n_{{\rm H}}$} &          
\colhead{$T$} &                    
\colhead{$R$} &                    
\colhead{$\Delta R$} &             
\colhead{$\eta _{{\rm peak}}$} &   
\colhead{$M _{{\rm shell}}$} &     
\colhead{$M _{{\rm Jeans}}$}       
\\
(${\rm cm^{-3}}$) & ($10^{51}$ erg)  & (Myr) & (cm$^{-3}$) & 
(K) & (pc)  & (pc) &    & ($M_{\odot}$) & ($M_{\odot}$) 
}
\startdata
     1  &     1.0 &    10.23 & 2.9E+00 &   568.6 &   115.9 &   22.84 &     3.0 & 3.0E+05 & 4.8E+05  \\
        &    10.0 &    10.84 & 6.7E+00 &   466.8 &   196.6 &   12.89 &     8.7 & 1.3E+06 & 2.4E+05  \\
        &    30.0 &    10.35 & 1.1E+01 &   419.7 &   250.9 &    9.17 &    14.4 & 2.6E+06 & 1.6E+05  \\
\hline
    10  &     1.0 &     3.39 & 5.9E+01 &   300.2 &    45.6 &    3.53 &     6.9 & 1.7E+05 & 4.1E+04  \\
        &    10.0 &     2.99 & 1.7E+02 &   239.8 &    75.2 &    1.72 &    21.5 & 6.7E+05 & 1.7E+04  \\
        &    30.0 &     2.63 & 2.4E+02 &   215.2 &    93.9 &    1.42 &    36.5 & 1.3E+06 & 1.2E+04  \\
\hline
   100  &     1.0 &     1.06 & 1.1E+03 &   191.6 &    17.9 &    0.68 &    14.2 & 9.3E+04 & 4.9E+03  \\
        &    10.0 &     0.83 & 2.4E+03 &   182.8 &    28.5 &    0.43 &    38.5 & 3.5E+05 & 3.0E+03  \\
        &    30.0 &     0.72 & 3.8E+03 &   186.4 &    35.2 &    0.33 &    58.4 & 6.4E+05 & 2.5E+03  \\
\hline
  1000  &     1.0 &     0.32 & 1.4E+04 &   202.5 &     6.9 &    0.19 &    19.5 & 5.1E+04 & 1.5E+03  \\
        &    10.0 &     0.25 & 3.1E+04 &   206.5 &    10.7 &    0.12 &    48.8 & 1.8E+05 & 1.0E+03  \\
        &    30.0 &     0.22 & 4.0E+04 &   223.0 &    13.2 &    0.12 &    69.8 & 3.4E+05 & 1.0E+03
\enddata
\tablecomments{
$Z$: metallicity of the ambient material,
$n_0$: ambient gas density,
$E_{{\rm SN}}$: explosion energy,
$t_{{\rm inst}}$: time when the gravitational instability first appears,
$n_{{\rm H}}$: volume-averaged Hydrogen number density in the shell,
$T$: mass-weighted average of temperature in the shell, 
$R$: radius of the shock front,
$\Delta R$: shell thickness,
$\eta _{{\rm peak}}$: peak wavenumber of the perturbations,
$M_{{\rm shell}}$: mass of the shell, and
$M_{{\rm Jeans}}$: Jeans mass, 
 at $t=t_{{\rm inst}}$}
\end{deluxetable*}

\subsection{Evolution of Fragments}

\begin{figure}[ht]
\epsscale{1.20}
\plotone{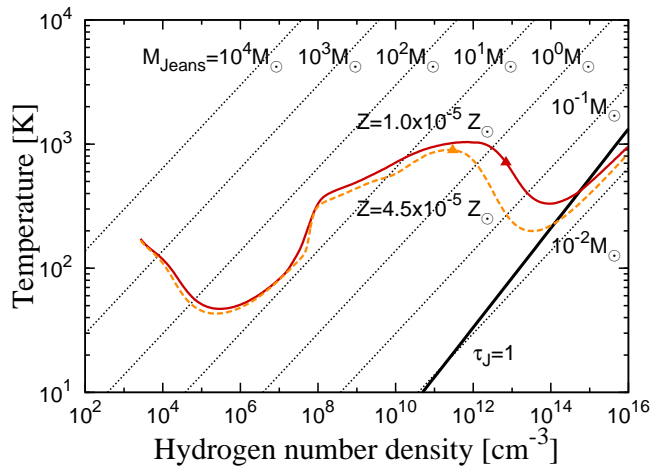}
\caption{
We plot the 
thermal evolution of the core of a collapsing fragment
for $(E_{{\rm SN}},n_0)
=(1\times 10^{52} \ {\rm erg} ,100 \ {\rm cm^{-3}})$.
Red and orange curves denote the trajectories 
for $Z=1.0\times 10^{-5} \ Z_{\odot }$ and 
$4.5\times 10^{-5} \ Z_{\odot }$, respectively.
The initial condition of the one-zone calculation
is determined using the result of our 
one-dimensional calculation.
The initial density and temperature are set as
the averaged values over the shell at the time
when the shell fragments.
Triangles are plotted at the points where 
the rapid dust cooling becomes dominant
[equation (\ref{dustdom:criterion}) is satisfied]
for each metallicity.
Thick solid line indicates the opacity limit characterized
by $\tau_{\rm J}$ = 1, where the cloud core
is optically thick to continuum radiation 
\citep[see {\it e.g.}][]{Omukai00}.
The instability that induces deformation of the fragment
occurs when the temperature decreases by rapid cooling
($\gamma _{{\rm eff}} \lesssim 1$).
The fragment continues to deform, inducing further fragmentation
until the temperature stops to decrease.
The corresponding Jeans mass at the point is
0.01--0.1 $M_{\odot}$ which is regarded as the characteristic
mass of a protostar.
}
\label{fig:nT}
\end{figure}

We investigate the subsequent evolution of the fragment 
for the models with large ambient gas density $n_0>10 \ {\rm cm^{-3}}$
where the SN-driven shell becomes gravitationally
unstable well before $10^7$ yr.
In these models, the time, $t_{{\rm inst}}$ when SN shells fragment is comparable to 
the free-fall timescale, $t_{{\rm ff}}$, of the shell at the time $t_{{\rm inst}}$.
The fragments are expected to contract gravitationally on the free-fall time scale
(equation \ref{one-zone:drhodt}).
Curves in Figure \ref{fig:nT} 
denote the evolutionary trajectories of 
the core of each fragment calculated by our one-zone code
for the metallicities 
$1.0\times 10^{-5} \ Z_{\odot }$ and $4.5\times 10^{-5} \ Z_{\odot }$
when SN explosion energy and ambient density
are $(E_{{\rm SN}},n_0) = (1\times 10^{52} \ {\rm erg},100 \ {\rm cm^{-3}})$.
The initial density, temperature and the chemical 
species' abundances are taken from the result of our one-dimensional 
calculation. We use the result at the time when the gravitational 
instability is triggered in the shell.

The temperature stays between a few hundred to one
thousand Kelvins
while the density increases over many orders of magnitudes.
The compressional heating is largely
balanced by radiative cooling until
the Hydrogen atomic number density increases to 
$n_{\rm H} \sim 10^{16} {\rm cm}^{-3}$.
There are, however, two slight declines
as clearly seen in Figure \ref{fig:nT}.
The first decline appears at
$n_{{\rm H}} \sim 10^4 \ {\rm cm^{-3}}$ 
where cooling by HD molecular lines is efficient.
The second decline appears at
$n_{{\rm H}} \sim 10^{13} \ {\rm cm^{-3}}$ 
where cooling by dust thermal emission is efficient.
In the latter regime, the gas cloud core
becomes further unstable and is likely to fragment
into multiple clumps \citep{Omukai05}.

The gravitational instabilities occur when
the dust cooling timescale is shorter than the free-fall timescale.
The gas cooling by dust is 
due to heat transfer from gas particles
to cold dust grains \citep[see][\S 7.4.3]{Stahler05}. 
Then, essentially the gas pressure determines the 
juncture for efficient dust cooling as
\begin{equation}
p
=n_{{\rm H}}kT
>
\frac{32G}{9\pi}
\left(
 \frac{a_d \rho _d}{Z_d \alpha }
\right)
^2 ,
\end{equation}
where 
$a_d$ and $\rho _d$ denote the mean radius 
and mass density of an individual dust grain, 
$Z_d$ is dust mass fraction, and
$\alpha $ is the accommodation coefficient for heat transfer
between gas and dust.
From the typical values of $a_d \sim 10^{-5} \ {\rm cm}$,
$\rho _d \simeq 3 \ {\rm g \ cm^{-3}}$ \citep{Love94},
$Z_d \sim 10^{-2}(Z/Z_{\odot })$,
and $\alpha \sim 0.1$, the region on the $n$-$T$ plane where 
dust cooling becomes efficient is
\begin{equation}
\frac{p}{k} = n_{{\rm H}}T
\gtrsim
4.9\times 10^{15} \ {\rm cm^{-3} \ K}
\left(
  \frac{Z/Z_{\odot}}{10^{-5}}
\right)
^{-2} .
\label{dustdom:criterion}
\end{equation}
The triangles in Figure \ref{fig:nT} denotes
the points where the gas clouds satisfy the condition
given by equation
(\ref{dustdom:criterion}).
The corresponding Jeans mass at the point is
$0.4 \ M_{\odot }$ 
for $Z=1.0 \times 10^{-5} \ Z_{\odot }$  and
$3 \ M_{\odot }$
for $4.5 \times 10^{-5} \ Z_{\odot }$.
The Jeans mass further decreases as the gas temperature 
rapidly decreases by dust cooling, as can be seen in Figure \ref{fig:nT}.

The rapid dust cooling
causes deformation of the cloud core into
a filamentary structure.
\citet{TsuribeOmukai08} show, using three-dimensional simulations,
that an initially perturbed gas cloud is significantly elongated
to filamentary shape 
and further fragments into small pieces
when dust cooling operates
for the metallicities
$10^{-6} \lesssim Z/Z_{\odot } \lesssim 10^{-5}$.
The deformation instability occurs 
when the deformation parameter, defined as 
the effective ratio of specific heats,
$\gamma _{{\rm eff}} \equiv d \log p / d \log \rho $,
is less than a critical value $\gamma _{{\rm crit}} = 1.097$
\citep{HanawaMatsumoto00,Lai00}. 
In our calculations,
$\gamma _{{\rm eff}}$
declines rapidly below the critical value
when the dust cooling becomes efficient.
The deformation parameter is 0.5 at a minimum 
when $n_{{\rm H}} \sim 10^{13} \ {{\rm cm^{-3}}}$
for $Z=1.0\times 10^{-5} \ Z_{\odot}$.
The value is common with
all explosion energies we set.
For $Z=4.5\times 10^{-5} \ Z_{\odot}$,
the minimum value of $\gamma _{{\rm eff}}$ is
0.3--0.4 at $n_{{\rm H}} \sim 10^{12} \ {\rm cm^{-3}}$.
The contracting gas clouds are strongly unstable to 
deformation in these regimes.

When the instability occurs, regardless of 
the SN explosion energies and metallicities,
the Jeans mass is 0.01--0.1 $M_{\odot}$ 
which is the characteristic mass of a fragment (protostar).
The ambient gas quickly accretes onto the protostars 
and their mass likely become larger \citep{Larson69}.
However, the subsequent evolution of the 
protostars will be highly dynamical.
According to \citet{PortegiesZwart10},
some of the protostars can attain velocities 
above the escape velocity of the parent gas cloud
through dynamical interactions among them.
They would then escape from the gas cloud
before the protostars become massive.
One natural outcome would then be the formation
of low-mass stars.


\section{DISCUSSION}

We have studied the evolution of an early
SN remnant with a metallicity of $\sim 10^{-5} \ Z_{\odot}$.
The supernova shell 
satisfies the criterion
for gravitational instability
at $t \simeq 10^5$--$10^6$ yr, to form
multiple fragments with the characteristic Jeans mass of 
$M_{{\rm Jeans}} \sim 10^3$--$10^4 \ M_{\odot}$
for a wide range of the SN
explosion energies, $E_{{\rm SN}} = 1\times 10^{51}$--$3\times 10^{52}$ erg,
and large ambient gas density, $n_0 \gtrsim 10 \ {\rm cm^{-3}}$.
We have then followed the thermal evolution
of one of the fragments using one-zone
calculations.
The gas cloud core evolves roughly
isothermally until the density increases 
to $n_{\rm H} \sim 10^{13} \ {\rm cm}^{-3}$,
where efficient cooling by dust thermal emission
brings the gas temperature from $\sim 10^3$ K to $\sim 300$ K rapidly.
The unstable core is expected to 
deform and fragment again into multiple clumps (protostars)
with masses of $\sim 0.01$--$0.1 \ M_{\odot }$
regardless the SN explosion energies and metallicities.
If some of the protostars escape from the host gas cloud
before growing in mass over $\sim 0.8 \ M_{\odot }$,
they would live long to the present day.
We argue that the shell fragmentation of
an early SN remnant is a viable scenario
for the formation of low-mass and extremely low-metallicity
stars.

There have been several theoretical studies
which suggest the formation of low-mass
stars in low-metallicity gas 
using numerical simulations.
For metallicites of $\sim 10^{-5}Z_{\odot}$, 
small mass clumps are formed
from a rotating gas cloud \citep{TsuribeOmukai08} and
in turbulent gas \citep{Dopcke11}
in many conditions these authors examined. 
Our hydrodynamic calculations are based on 
SN explosions of an early generation of stars. 
The end point of the hydrodynamic simulation of the SN remnant, 
when the shocked shell
becomes gravitationally unstable, provides
physically motivated initial conditions for our
one-zone calculations.
Therefore, we are able to propose, for the first time, 
a consistent model of how low-metallicity stars are formed 
in the early universe. 

We note that there are still limitations in the one-dimensional and 
one-zone simulations.
Non-spherical structures of an SN remnant
might have substantial effects on the shell fragmentation.
\citet{Chevalier75} and \citet{ChevalierImamura82} 
show that shell instabilities might 
enhance density perturbations around an SN remnant 
to form dense clumps in early epochs of the remnant evolution.
Dense gas clumps could form even earlier than we have shown 
in this paper.
Clearly, full three-dimensional calculations are needed to study in
detail the evolution of an SN shell.
Our calculations may provide basic configurations for such studies on
the formation of low-mass stars
in low-metallicity gas.

Currently standard models of galaxy formation
are based on the hierarchical growth of
cold dark matter halos. 
Recent high-resolution $N$-body simulations
show that tens of thousands of early small halos
assemble a Milky Way size halo \citep{Gao10,Wang11}.
Although stars formed in the earliest halos 
are likely populated in the central bulge region,
there should be a substantial fraction of the remnants
of early halos orbiting in the halo of the Galaxy
\citep{Tumlinson10,Gao10}.
\citet{Caffau11} may have found one of such ``fossils''.
Future surveys of metal-poor stars
will provide the relative number fraction
of extremely metal-poor stars, which will
then constrain the overall formation efficiency
of low-mass stars in the scenario we propose here.

\acknowledgments

The authors thank 
K. Omukai, 
M. N. Machida,
K. Iwasaki,
T. Tsuribe,
and colleagues at IPMU
for fruitful discussions and helpful advice.
We are grateful to an anonymous referee of ApJ,
who suggested us to derive the metallicity dependence
of the critical pressure for the dust cooling.
Part of the simulations were performed at the Center
for Computational Astrophysics at 
the National Astronomical Observatory of Japan.
This work is supported in part by the Grants-in-Aid for Young Scientists
(S: 20674003, B: 21740139) by the Japan Society for the Promotion of
Science. N.Y. acknowledges financial support from World Premier
International Research Center Initiative (WPI Initiative), MEXT, Japan.


\begin{thebibliography}{}

\bibitem[Alvarez et al.(2006)]{Alvarez06} Alvarez, M.~A., Bromm, 
V., \& Shapiro, P.~R.\ 2006, \apj, 639, 621 
\bibitem[Aver et al.(2010)]{Aver10} Aver, E., Olive, K.~A., 
\& Skillman, E.~D.\ 2010, \jcap, 5, 3 

\bibitem[Beers 
\& Christlieb(2005)]{Beers05} Beers, T.~C., \& Christlieb, N.\ 2005, \araa, 43, 531 
\bibitem[Bromm et al.(2003)]{Bromm03} Bromm, V., Yoshida, N., 
\& Hernquist, L.\ 2003, \apjl, 596, L135 
\bibitem[Bromm et al.(2009)]{Bromm09} Bromm, V., Yoshida, N., 
Hernquist, L., \& McKee, C.~F.\ 2009, \nat, 459, 49 
\bibitem[Bromm 
\& Yoshida(2011)]{BrommYoshida11} Bromm, V., \& Yoshida, N.\ 2011, \araa, 49, 373 

\bibitem[Caffau et al.(2011)]{Caffau11} Caffau, E., Bonifacio, 
P., Fran{\c c}ois, P., et al.\ 2011, \nat, 477, 67 
\bibitem[Chevalier 
\& Theys(1975)]{Chevalier75} Chevalier, R.~A., \& Theys, J.~C.\ 1975, \apj, 195, 53
\bibitem[Chevalier 
\& Imamura(1982)]{ChevalierImamura82} Chevalier, R.~A., \& Imamura, J.~N.\ 1982, \apj, 261, 543 

\bibitem[Dopcke et al.(2011)]{Dopcke11} Dopcke, G., Glover, 
S.~C.~O., Clark, P.~C., \& Klessen, R.~S.\ 2011, \apjl, 729, L3 

\bibitem[Elmegreen(1994)]{Elmegreen94} Elmegreen, B.~G.\ 1994, 
\apj, 427, 384

\bibitem[Flower et al.(2000)]{Flower00} 
Flower, D. R., Le Bourlot, J., Pineau des Forets, G., \& Roueff, E. \ 2000, 
\mnras, 314, 753

\bibitem[Galli \& Palla (1998)]{GalliPalla98}
Galli D., \&  Palla, F. \ 1998, \aap, 335, 403
\bibitem[Gao et al.(2010)]{Gao10} Gao, L., Theuns, T., Frenk, 
C.~S., et al.\ 2010, \mnras, 403, 1283 
\bibitem[Greif et al.(2010)]{Greif10} Greif, T.~H., Glover, 
S.~C.~O., Bromm, V., \& Klessen, R.~S.\ 2010, \apj, 716, 510 

 
\bibitem[Hanawa 
\& Matsumoto(2000)]{HanawaMatsumoto00} Hanawa, T., \& Matsumoto, T.\ 2000, \pasj, 52, 241 
\bibitem[Hosokawa et al.(2011)]{Hosokawa11} Hosokawa, T., Omukai, 
K., Yoshida, N., \& Yorke, H.~W.\ 2011, Science, 334, 1250 


\bibitem[Iocco et al.(2009)]{Iocco09} Iocco, F., Mangano, G., 
Miele, G., Pisanti, O., \& Serpico, P.~D.\ 2009, \physrep, 472, 1
\bibitem[Iwasaki 
\& Tsuribe(2008)]{IwasakiTsuribe08} Iwasaki, K., \& Tsuribe, T.\ 2008, \pasj, 60, 125 
\bibitem[Iwasaki et al.(2011a)]{IIT11i} Iwasaki, K., Inutsuka, 
S.~i., \& Tsuribe, T.\ 2011a, \apj, 733, 17
\bibitem[Iwasaki et al.(2011b)]{IIT11ii} Iwasaki, K., Inutsuka, 
S.~i., \& Tsuribe, T.\ 2011b, \apj, 733, 16

\bibitem[Kitayama et al.(2004)]{Kitayama04} 
Kitayama, T., Yoshida, N., Susa, H., Umemura, M.
\ 2004, \apj, 613, 631
\bibitem[Kitayama 
\& Yoshida(2005)]{KitayamaYoshida05} Kitayama, T., \& Yoshida, N.\ 2005, \apj, 630, 675

\bibitem[Lai(2000)]{Lai00} Lai, D.\ 2000, \apj, 540, 946 
\bibitem[Larson et al.(2011)]{Larson11} Larson, D., Dunkley, J., 
Hinshaw, G., et al.\ 2011, \apjs, 192, 16 
\bibitem[Larson(1969)]{Larson69} Larson, R.~B.\ 1969, \mnras, 
145, 271 
\bibitem[Lipovka et al.(2005)]{Lipovka05} 
Lipovka, A., Nunez-Lopez, R., \& Avila-Reese, V. \ 2005, \mnras, 361, 850
\bibitem[Love et al.(1994)]{Love94} Love, S.~G., Joswiak, 
D.~J., \& Brownlee, D.~E.\ 1994, \icarus, 111, 227 

\bibitem[Machida et al.(2005)]{Machida05} Machida, M.~N., 
Tomisaka, K., Nakamura, F., \& Fujimoto, M.~Y.\ 2005, \apj, 622, 39 


\bibitem[Nagakura et al.(2009)]{Nagakura09} Nagakura, T., 
Hosokawa, T., \& Omukai, K.\ 2009, \mnras, 399, 2183 

\bibitem[Omukai(2000)]{Omukai00} Omukai, K.\ 2000, \apj, 534, 
809 
\bibitem[Omukai 
\& Yoshii(2003)]{OmukaiYoshii03} Omukai, K., \& Yoshii, Y.\ 2003, \apj, 599, 746 
\bibitem[Omukai et al.(2005)]{Omukai05} Omukai, K., Tsuribe, T., 
Schneider, R., \& Ferrara, A.\ 2005, \apj, 626, 627 
\bibitem[Omukai 
\& Nishi(1998)]{OmukaiNishi98} Omukai, K., \& Nishi, R.\ 1998, \apj, 508, 141 
\bibitem[Omukai et al.(2010)]{OmukaiHosokawaYoshida10} Omukai, K., Hosokawa, 
T., \& Yoshida, N.\ 2010, \apj, 722, 1793
\bibitem[Ostriker 
\& McKee(1988)]{OstrikerMcKee88} Ostriker, J.~P., \& McKee, C.~F.\ 1988, Reviews of Modern Physics, 60, 1 


\bibitem[Pollack et al.(1994)]{Pollack94} Pollack, J.~B., 
Hollenbach, D., Beckwith, S., et al.\ 1994, \apj, 421, 615 
\bibitem[Portegies Zwart et 
al.(2010)]{PortegiesZwart10} Portegies Zwart, S.~F., McMillan, S.~L.~W., \& Gieles, M.\ 2010, \araa, 48, 431 



\bibitem[Santoro 
\& Shull(2006)]{SantoroShull06} Santoro, F., \& Shull, J.~M.\ 2006, \apj, 643, 26 
\bibitem[Spitzer(1978)]{Spitzer78} Spitzer, L., Jr.\ 1978, 
\jrasc, 72, 349 
\bibitem[Schneider et al.(2003)]{Schneider03} Schneider, R., 
Ferrara, A., Salvaterra, R., Omukai, K., \& Bromm, V.\ 2003, \nat, 422, 869
\bibitem[Schneider et al.(2012)]{Schneider12} Schneider, R., 
Omukai, K., Bianchi, S., \& Valiante, R.\ 2012, \mnras, 419, 1566 
\bibitem[Sedov(1958)]{Sedov58} Sedov, L.~I.\ 1958, Reviews of 
Modern Physics, 30, 1077
\bibitem[Shu et al.(2002)]{Shu02} Shu, F.~H., Lizano, S., 
Galli, D., Cant{\'o}, J., \& Laughlin, G.\ 2002, \apj, 580, 969 
\bibitem[Stahler 
\& Palla(2005)]{Stahler05} Stahler, S.~W., \& Palla, F.\ 2005, The Formation of Stars, by Steven W.~Stahler, Francesco Palla, pp.~865.~ISBN 3-527-40559-3.~Wiley-VCH , January 2005.,  
\bibitem[Sutherland 
\& Dopita(1993)]{SutherlandDopita93} Sutherland, R.~S., \& Dopita, M.~A.\ 1993, \apjs, 88, 253

\bibitem[Tsuribe 
\& Omukai(2008)]{TsuribeOmukai08} Tsuribe, T., \& Omukai, K.\ 2008, \apjl, 676, L45 
\bibitem[Tumlinson(2010)]{Tumlinson10} Tumlinson, J.\ 2010, \apj, 
708, 1398 

\bibitem[Whalen et al.(2004)]{Whalen04} Whalen, D., Abel, T., 
\& Norman, M.~L.\ 2004, \apj, 610, 14 
\bibitem[Whalen et al.(2008)]{Whalen08} Whalen, D., van Veelen, 
B., O'Shea, B.~W., \& Norman, M.~L.\ 2008, \apj, 682, 49 
\bibitem[Wang et al.(2011)]{Wang11} Wang, H., Mo, H.~J., Jing, 
Y.~P., Yang, X., \& Wang, Y.\ 2011, \mnras, 413, 1973 
\bibitem[Whalen et al.(2010)]{Whalen10} Whalen, D., Hueckstaedt, 
R.~M., \& McConkie, T.~O.\ 2010, \apj, 712, 101 

\bibitem[Yoshida et al.(2003)]{Yoshida03} Yoshida, N.,
Abel, T., Hernquist, L., Sugiyama, N. \ 2003, \apj, 592, 645
\bibitem[Yoshida et al.(2006)]{Yoshida06} Yoshida, N., Omukai, 
K., Hernquist, L., \& Abel, T.\ 2006, \apj, 652, 6 
\bibitem[Yoshida et al.(2008)]{Yoshida08} Yoshida, N., Omukai, K.,
Hernquist, L. \ 2008, Science, 321, 669

\end{thebibliography}
\end{document}